\documentclass[twocolumn,showpacs,preprintnumbers,amsmath,amssymb]{revtex4} 
\usepackage{dcolumn}
\usepackage{graphicx,color,epsfig}
\def\bc{\begin{center}}
\def\ec{\end{center}}
\def\be{\begin{eqnarray}}
\def\ee{\end{eqnarray}}

\begin{document}

\title{Perturbative QCD for $J/\psi$ Inclusive Production Via Initial State Radiation at $e^+e^-$ collider}
\author{Bin Gong}
\email{twain@ihep.ac.cn}
\author{Yu-Dong Wang}
\email{wangyudong@ihep.ac.cn}
\author{Jian-Xiong Wang}
\email{jxwang@ihep.ac.cn}
\affiliation{%
Institute of High Energy Physics, Chinese Academy of Sciences, Beijing 100049, China\\
School of Physics, University of Chinese Academy of Sciences, Beijing 100049, China
}

\begin{abstract}
Up to the next-leading order (NLO) of quantum chromodynamics (QCD), 
the process $e^+e^-\to J/\psi +X$ with the center-of-mass (CM) energy range from 3.7 to 10.6 GeV is calculated.
At 10.6 GeV, the results is consistent with the experiment results at the Belle.
However,  the predictions are much smaller than the measurement at  BESIII at low CM energy range from 3.7 to 4.6 GeV. 
This indicates that the convergence of  QCD perturbative expansion becomes worse as the CM energy becomes lower and closer to the  
inclusive $J/\psi$ production threshold. 
For a further study of  the QCD mechanism on $J/\psi$ production at $e^+e^-$ collider with different CM energy,
the initial state radiation effect of $e^+e^-\to J/\psi+gg$ and $e^+e^-\to J/\psi+c \bar{c}$  are calculated at the  QCD NLO. 
The results are plotted and the numbers of events for different CM energy bins are provided for  the designed  SuperKEKB. 
This provides a method to precisely test the validity of perturbative prediction on $J/\psi$ production in future measurements.
\end{abstract}

\maketitle

\section{Introduction} 

QCD is the theory of strong interaction between quarks and gluons 
in the Standard Model.  It exhibits two main properties, color confinement and asymptotic freedom\cite{Gross:1973id, Politzer:1973fx}.
Therefore, there are both perturbative and non-perturbative parts of QCD in the calculation of processes involves hadrons. 
The study on $J/\psi$ related processes provides a good method to probe both perturbative and non-perturbative aspects of QCD dynamics.  
On one side, $J/\psi$'s leptonic decays make it easy to be measured in  the experiments. 
On the other side, $J/\psi$ is a bound state of $c\bar c$ pair where $c$ quark is heavy so that $J/\psi$ related processes can 
be well factorized into perturbative and non-perturbative parts in theoretical calculation.  
In 1995, in order to explain the experimental measurements on $J/\psi$ and $\psi^\prime$ production at  the Tevatron\cite{Abe:1992ww}, 
a non-relativistic QCD (NRQCD) factorization formalism was proposed based on the color-octet (CO) mechanism\cite{Bodwin:1994jh}. 
It allows consistent theoretical prediction to be made and improved order by order in the strong coupling constant $\alpha_s$  and the
heavy quark relative velocity $v$, although the CO NRQCD long-distance matrix elements (LDMEs) which thought to be universal 
can only be obtained by fitting experimental data.

In the last twenty years, many important progresses have be achieved on both experimental and theoretical studies for $J/\psi$ related processes at different colliders. 
There are very precise experimental measurements on $J/\psi$ production and polarization at the LHC\cite{Khachatryan:2010yr,Aaij:2013nlm,Aaij:2014qea}, with their theoretical predictions
extending to NLO
\cite{Campbell:2007ws,Gong:2008sn,Gong:2008ft,Butenschoen:2010rq,Ma:2010yw,Butenschoen:2011yh,Ma:2010jj,Butenschoen:2012px,
Chao:2012iv,Gong:2012ug,Feng:2018ukp}.
	
But things are quite different in $e^+e^-$ colliders. 
The signature for CO production of $J/\psi$ in $e^{+} e^{-}$ annihilation at B-factories was suggested in Ref.~\cite{Braaten:1995ez} 
and its contribution is at the endpoint of $J/\psi$ momentum spectra due to the kinematics of the two-body final states.
The cross sections for inclusive $J/\psi$ production in $e^+e^-$ annihilation were measured by BABAR and 
the Belle\cite{Aubert:2001pd,Abe:2001za,Abe:2002rb,Aubert:2005tj}. But this CO signal was not observed.  
In Ref.~\cite{Fleming:2003gt}, the authors tried to spread the CO signal by resuming the CO contribution. 
There was a suggestion in Ref.~\cite{Wang:2003fw} to observe the CO contribution at $e^+ e^- \rightarrow J/\psi +\gamma+X$ 
where CO channels have larger contribution than the color-singlet (CS) channels
at different range in energy spectrum of the photon . 
The experimental results are several times larger than  the leading-order CS prediction\cite{Keung:1980ev,Liu:2002wq}. 
This large discrepancy was resolved by including  the NLO QCD corrections, relativistic corrections, and feed-down contribution from higher excited states 
(see e.g. Refs.~\cite{Zhang:2006ay,Ma:2008gq,Gong:2009kp,Gong:2009ng}). 
Therefore the contributions from CS channels can already explain the experimental data and almost no space is left for CO contribution~\cite{Zhang:2009ym}. 

In experimental measurements, as many exotic states will decay into $J/\psi$.  $J/\psi$ production is very important background in the search for those exotic states. The precise measurement of $e^+e^-\rightarrow J/\psi \pi^+\pi^-$ was performed by BESIII Collaboration\cite{Ablikim:2016qzw}.  
Except the contributions from decay of exotic states, the contribution for $J/\psi$ inclusive production from continuous background obtained in this measurement is much larger. It may provides space for CO contribution~\cite{Li:2014fya}. 
Thus a detailed study for inclusive $J/\psi$ production in $e^+e^-$ collider with  the center-of-mass (CM) energy range from 3.7 to 4.6 GeV
is interesting. But the energy range is near inclusive  $J/\psi$ production threshold, so the validity of perturbative QCD calculation is strongly doubted. 
Previous studies have already shown that the CS  result at the NLO can describe the  the experimental measurement on inclusive  $J/\psi$ production at B-factories energy (10.58 GeV). Therefore the question, from which CM energy can CS perturbative results describe inclusive  $J/\psi$ production 
at $e^+e^-$ collider, is the point we will address in this work.

This paper is organized as follows. we give a detailed study for inclusive $J/\psi$ production in $e^+e^-$ collider with CM energy range 
from 3.7 to 10.58 GeV in Sec. II.  In Sec. III, we suggest to measure inclusive $J/\psi$ production by using the initial state radiation (ISR) effect at the B-factories and  
present a detailed calculation at the  QCD NLO for it.  The summary and conclusion are in Sec. IV.

%*****************************************************************************************
%*******************************SECTION 2************************************************
%*****************************************************************************************

\section{The cross section at the NLO from 3.70 to 10.58 GeV}
As the CM energy is not enough for $J/\psi+c\bar{c}$ production at the BESIII, the calculation is almost same as $J/\psi+gg$ case\cite{Gong:2009kp} at  B factories, but with much smaller $\sqrt{s}$. To perform the calculation, FDC package\cite{FDC} is used to generate quad-precision FORTRAN Codes, which is essential 
to deal with serious numerical unstable problem in the calculation of virtual corrections near threshold. 
A two-cutoff method\cite{Harris:2001sx} is used to treat the infrared divergences in real correction processes, and after the check of cutoff independence, $\delta_s=10^{-3}$ and $\delta_c=2\times 10^{-5}$  are chosen. More details about the calculations can be found in Refs.\cite{Gong:2009kp,Gong:2009ng}.

For numerical results, the approximation  $M_{J/\psi}=2m_c$ is our default choice. 
$m_c$ is set to be 1.4, 1.5 and 1.6 GeV and the renormalization scale $\mu_r$ is chosen as 2$m_c$ or $\sqrt{s}/2$ 
in the calculations which give an uncertainty of the results. 
The radial wave function at the origin of $J/\psi$, $|R_s(0)|^2$,  is set to $1.006$ GeV$^3$ for $m_c=1.5$ GeV. $m_\pi$ is set to 139.6 MeV according to PDG\cite{Patrignani:2016xqp}. For the value of strong coupling constant $\alpha_s$, two-loop formula is used. To produce two $\pi$ in the final states, a cut, $m^2_X\ge (2m_{\pi})^2$ is introduced on the 
invariant mass of $gg$ for $e^+e^-\to J/\psi gg$, $ggg$ for $e^+e^-\to J/\psi ggg$ and  $g q\bar{q}$ for $e^+e^-\to J/\psi g q\bar{q}$. 

In Table \ref{tab1}, the total cross sections of inclusive $J/\psi$ production at the NLO with different $\sqrt{s}$ are shown, as well as the ratio of NLO  results to LO ones. Unlike the B-factory case, the ratio here are much larger, ranging from  1.90 to 2.73. In  some sense, this larger ratio indicates that the convergence of QCD perturbative expansion becomes bad here.   
\begin{table}[htp]
\begin{center}
\footnotesize
\begin{tabular*}{80mm}{cccc}
%{|c@{\extracolsep{\fill}}|c|c|c|}
\hline\hline 
$\sqrt{s}$(GeV)   & $\sigma^{(0)}$(pb)   & $\sigma^{(1)}$(pb)  & $\sigma^{(1)}/\sigma^{(0)}$\\
\hline
3.7 & 1.027$\pm$0.001 & 2.799 $\pm$ 0.003 & 2.73\\
3.8 & 1.168$\pm$0.001 & 2.911 $\pm$ 0.003 & 2.56\\
3.9 & 1.262$\pm$0.001 & 3.060 $\pm$ 0.003 & 2.42\\
4.0 & 1.321$\pm$0.001 & 3.054 $\pm$ 0.003 & 2.31\\
4.1 & 1.354$\pm$0.001 & 3.004 $\pm$ 0.003 & 2.22\\
4.2 & 1.368$\pm$0.001 & 2.924 $\pm$ 0.004 & 2.14\\
4.3 & 1.369$\pm$0.001 & 2.828 $\pm$ 0.004 & 2.07\\
4.4 & 1.359$\pm$0.001 & 2.723 $\pm$ 0.004 & 2.00\\
4.5 & 1.343$\pm$0.001 & 2.613 $\pm$ 0.004 & 1.95\\
4.6 & 1.321$\pm$0.001 & 2.506 $\pm$ 0.003 & 1.90\\
\hline\hline
\end{tabular*}
\hspace{1cm}
\caption{\label{tab1}Cross sections at different CM energy with renormalization scale $\mu$=2$m_c$=3.0GeV.}
\end{center}
\end{table}

\begin{figure}
\begin{center}
\includegraphics[width=8cm]{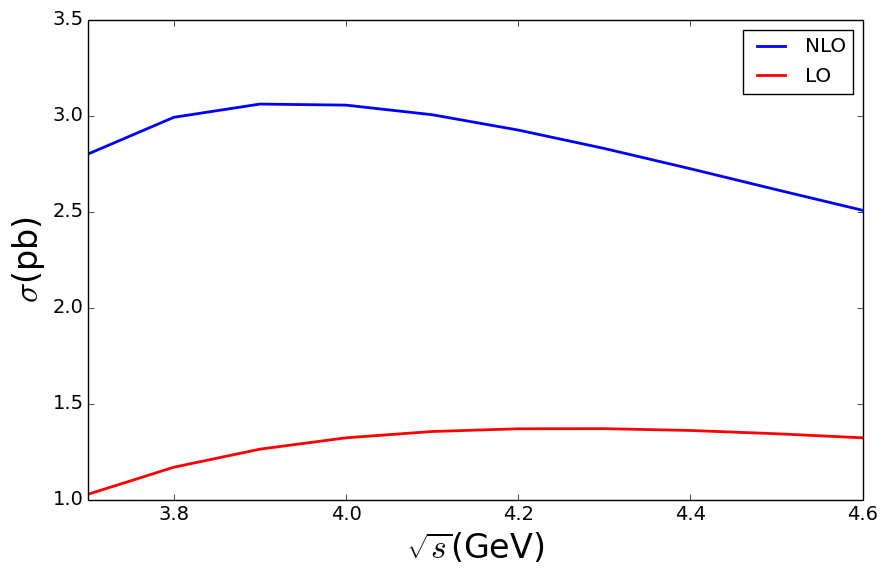}
\caption{\label{total1}  LO and NLO cross sections as functions of $\sqrt{s}$ } 
\end{center}
\end{figure}

\begin{figure}
\begin{center}
\includegraphics[width=8cm]{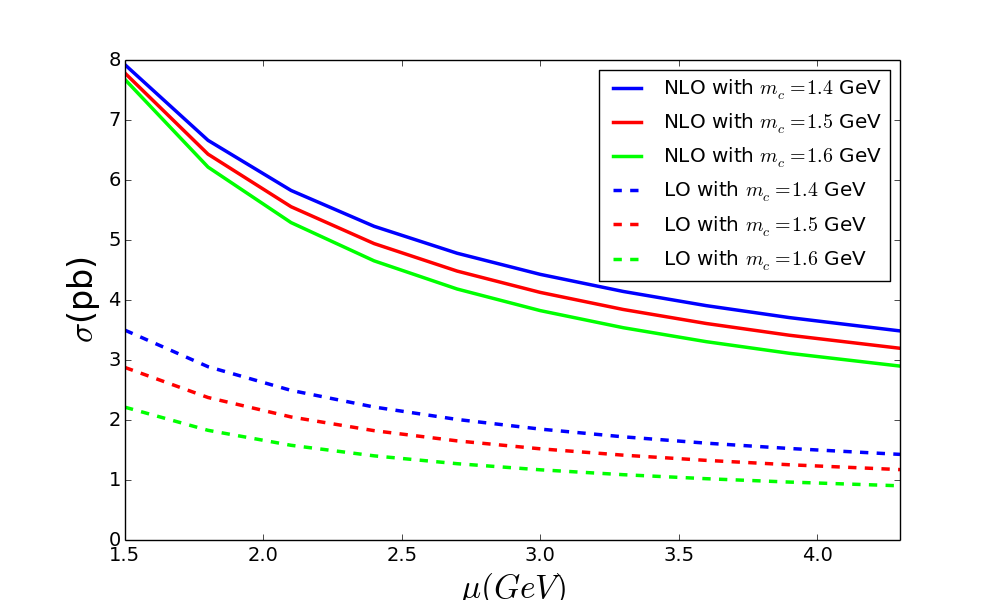}
\caption{\label{total2} Renormalization scale dependence of inclusive $J/\psi$ production at $\sqrt{s}$=4.0 GeV.}
\end{center}
\end{figure}

In Fig.\ref{total1}, total cross section for inclusive $J/\psi$ production is shown. The scale dependence of total cross section with $\sqrt{s}=4.0$ GeV is presented in Fig.~\ref{total2}.
It  shows clearly that the renormalization scale dependence is not improved for  the  NLO results in comparison with LO ones, 
and it gives more evidences that the convergence of QCD perturbative expansion become bad in this case.  

\begin{figure}
\begin{center}
\includegraphics[width=8cm]{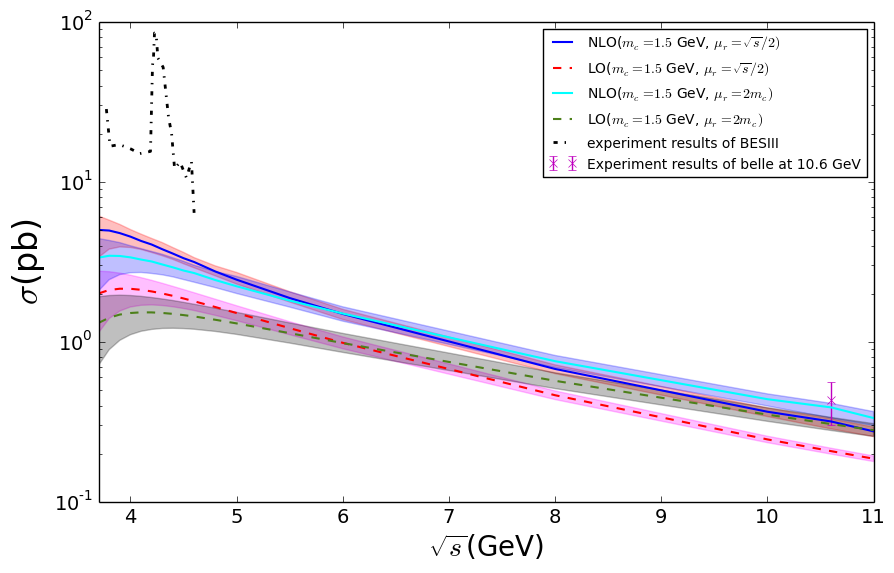}
\caption{\label{total3} LO and NLO theoretical prediction comparing with experimental data. }
\end{center}
\end{figure}

In Fig.~\ref{total3}, we compare the results with experimental measurements in Ref.\cite{Ablikim:2016qzw}. It can be seen that both LO and  the NLO theoretical predictions are far away from  experimental data, even though there is a large K factor.  It probably means that the perturbative calculation becomes very bad in this situation. 
As the CM energy becomes lower and closer to the  inclusive $J/\psi$ production threshold,  
the theoretical result of perturbative calculation loses its convergence gradually.
On the other side close to 10.58 GeV, the result is consistent with the measurements from Belle\cite{Pakhlov:2009nj}.
It is a interesting question to know the boundary where the QCD perturbative calculation is not suitable any more. 
Considering that the energy in the collider is limited and it is impossible to obtain an experimental curve like the one in  Fig.~\ref{total3} to  compare with ours.
However, there is another way in which we can do the comparison by utilizing the ISR effect of  QED.  The ISR effect of $e^+e^-\to J/\psi+X$ is discussed in next section. 
%*****************************************************************************************
%*******************************SECTION 3************************************************
%*****************************************************************************************
\section{The ISR contribution at B-factories}
In B-factories, there were heavy quarkonium related experimental measurement via ISR effect \cite{Yuan:2007sj,Wang:2007ea},
and there was also heavy quarkonium related theoretical study via ISR at LO \cite{Chang:2010am}.

In this section, the ISR  effect in the processes $e^+e^-\to J/\psi gg$  and $e^+e^-\to J/\psi c\bar{c}$ is numerically calculated at the SuperKEKB energy. 
With the ISR factorization formula from the factorization of the mass and  the infrared singularities\cite{Lee:1964is}, the total cross sections can be represented as:
\begin{equation}
\label{eq:4:1}
\begin{split}
\sigma(s)&=\iint dx_1dx_2 D_{e^-}(x_1,s)D_{e^+}(x_2,s)\sigma_R(x_1x_2s),
\end{split}
\end{equation}
where $D_{e^{-(+)}}(x,s)$ is the distribution function of the probability to find an electron (positron) with a momentum fraction $x$ within the original electron (positron). 
Without loss of generality, we use $D_e(x,s)$ instead in the following discussion. $\sigma_R(x_1x_2s)$ is the cross section with reduced CM energy $s^\prime=x_1x_2s$.
$D_e(x,s)$ satisfies following evolution equation (see as Ref.\cite{Kuraev:1985hb,Nicrosini:1986sm,Greco:2016izi})
\begin{equation}
\label{eq:evolution}
D(x,s)=\delta(1-x)+\int_{m_e^2}^s\frac{dQ^2}{Q^2}\frac{\alpha(Q^2)}{2\pi}\int_x^1\frac{dz}{z}P(z)D_e(\frac{x}{z},Q^2)
\end{equation}
with 
\begin{equation}
\alpha(Q^2)=\dfrac{\alpha}{1-(\alpha/3\pi)\mathrm{ln}(Q^2/m_e^2)}
\end{equation}
and
\begin{equation}
P(z)=\dfrac{1+z^2}{1-z}-\delta(1-z)\int_0^1 dx \dfrac{1+x^2}{1-x}
\end{equation}
is the regularized splitting function.
By defining $x=x_1x_2$, we have 
\begin{equation}
\label{eq:4:2}
\begin{split}
\sigma(s)&=\int dx F(x,s)\sigma_R(xs),
\end{split}
\end{equation}
with 
\begin{equation}
\label{eq:4:2x}
\begin{split}
F(x,s)&\equiv\int^1_x dx_1 \frac{1}{x_1} D_{e}(x_1,s)D_{e}(\frac{x}{x_1},s).
\end{split}
\end{equation}
On the other hand, if we define
\begin{equation}
\label{eq:4:x1x}
\begin{split}
R(x,s)&\equiv\int\limits^1_x dx_1 \int\limits^1_{x/x_1} dx_2  D_{e}(x_1,s)D_{e}(x_2,s),
%\\
 %        &=\int^1_x dt F(t,s).
\end{split}
\end{equation}
It can be obtained, with the substitution $x_2=t/x_1$,
\begin{equation}
\label{eq:4:x1}
\begin{split}
%R(x,s)&=\int^1_x dx_1 \int^{x_1}_x dt \frac{1}{x_1}  D_{e}(x_1,s)D_{e}(t/x_1,s), \\
  %       &=\int^1_x dt \int^1_t dx_1\frac{1}{x_1}  D_{e}(x_1,s)D_{e}(t/x_1,s), \\
R(x,s)&=\int^1_x dt F(t,s).
\end{split}
\end{equation}
%Meanwhile the cross section can also be represented as
%\begin{equation}
%\label{eq:4:x1}
%\begin{split}
%\sigma(s)&=R(x,s)\sigma_R(xs),\\
%R(x,s)&=\int^1_x dx_1 \int^1_{x/x_1} dx_2  D_{e^-}(x_1,s)D_{e^+}(x_2,s),\\
%         &=\int^1_x dt F(t,s).
%\end{split}
%\end{equation}
Therefore $F(x,s)$ can be obtained through taking the derivative of $R(x,s)$ (with negative sign).
It follows from Eq.(\ref{eq:evolution}) that $R(x,s)$ satisfies the equation 
\begin{equation}
\label{eq:4:x2}
\begin{split}
R(x,s)&=1+\int^s_{m_e^2} \frac{\alpha (Q^2)}{\pi}\frac{dQ^2}{Q^2} \int^1_x dzP(z)R(\frac{z}{x},Q^2).
\end{split}
\end{equation}
%
%where there are the running coupling constant $\alpha (Q^2)=\alpha/(1-\frac{\alpha}{3\pi}{\rm In}\frac{Q^2}{m^2})$ and spliting function $P(z)=\frac{1+z^2}{1-z}-\delta (1-z)\int^1_0 dx\frac{1+x^2}{1-x}$.
Defining another function 
\begin{equation}
G(x,s)\equiv \int^1_x dt D(t,s),
\end{equation}
$D(x,s)$  can be obtained through taking the derivative of $G(x,s)$.
Again from Eq.(\ref{eq:evolution}), a similar equation as Eq.(\ref{eq:4:x2}) is found for $G(x,s)$
\begin{equation}
\label{eq:4:x3}
\begin{split}
G(x,s)&=1+\int^s_{m_e^2}  \frac{\alpha (Q^2)}{2\pi}\frac{dQ^2}{Q^2} \int^1_x dzP(z)G(\frac{z}{x},Q^2).
\end{split}
\end{equation}
This equation differs from Eq.\eqref{eq:4:x2} through the substitution $\alpha (Q^2) \to \alpha(Q^2)/2$.
Following the procedures in Ref.\cite{Kuraev:1985wb}, $R(x,s)$ and $G(x,s)$ is obtained as 
\begin{equation}
\label{eq:4:x4}
\begin{split}
R(x,s)&=(1-x)^{\beta_l}(1+\frac{3}{4}\beta_l)+\frac{\beta_l}{2}(\frac{1}{2}x^2+x-\frac{3}{2}) +{\cal O}(\beta_l^2), \\
G(x,s)&=(1-x)^{\beta_l/2}(1+\frac{3}{8}\beta_l)+\frac{\beta_l}{4}(\frac{1}{2}x^2+x-\frac{3}{2}) +{\cal O}(\beta_l^2), \\
\end{split}
\end{equation}
with
\begin{equation}
\beta_l=\frac{2\alpha}{\pi}\left[\mathrm{log}\left(\dfrac{s}{m_e^2}\right) -1 \right].
\end{equation}
Thus $F(x,s)$ and $D_e(x,s)$ are obtained as 
\begin{equation}
\label{eq:4:3}
\begin{split}
F(x,s)&=\beta_l(1+\frac{3}{4}\beta_l)(1-x)^{\beta_l-1}-\frac{\beta_l}{2}(1+x)+O(\beta^2_l),\\
D_e(x,s)&=\frac{\beta_l}{2}(1+\frac{3}{8}\beta_l)(1-x)^{\beta_l/2-1}-\frac{\beta_l}{4}(1+x)+O(\beta^2_l),\\
\end{split}
\end{equation}

In  the SuperKEKB\cite{Akai:2018mbz,Lalwani:2018dgg}, it will collide electrons at 7 GeV with positrons at 4 GeV. The invariant variable $\sqrt{s}=\sqrt{112}$ GeV.
The half-crossing angle $\theta_c$ in the detector is 41.5 mrad, within which particles can not be measured.
The calculations are performed at the CM frame of partons, then the Lorentz boost is performed from this frame to the laboratory frame for all the involved particles.
Here we use a cut to make sure that the angle between outgoing $J/\psi$ and beam at the laboratory frame is larger than the cross angle, $\theta_{J/\psi}>\theta_c$.
The same cut is also applied for the recoiling particle $X$ (here the momentum of $X$ is the sum of all other final state particles), namely $\theta_X>\theta_c$.
In really experimental measurement, a hadron is reconstructed from its decay products. Thus even a hadron is inside the cross angle, it could be still observed if its decay products are not. The cut will work better when involves the related Monte Carlo simulation in data analysis.

\begin{figure}
\begin{center}
\includegraphics[width=8cm]{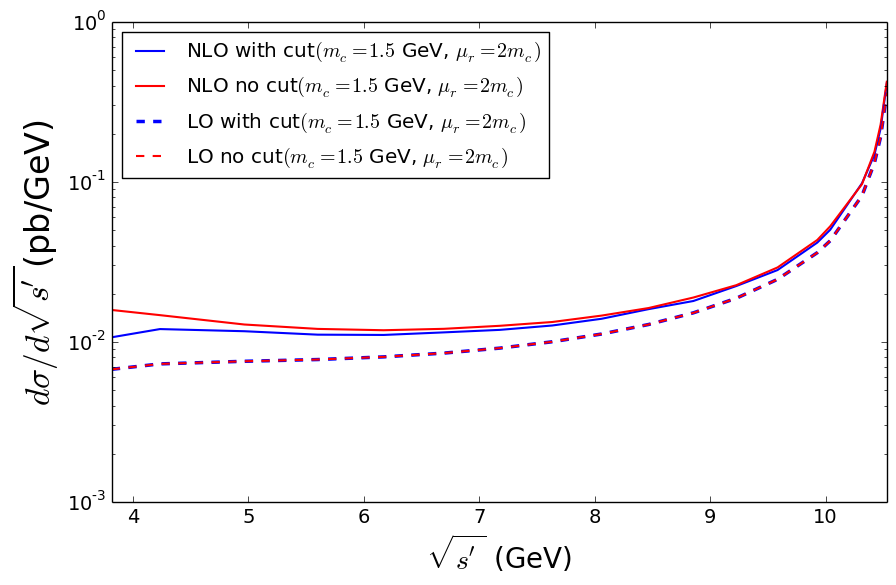}
\caption{\label{fig:ISR1} ISR effect for $J/\psi+gg$ at QCD LO and  NLO with and without cut.}  
\end{center}
\end{figure}

In  Fig.~\ref{fig:ISR1}, the ISR results with reduced CM energy ranging from 3.8  to 10.58 GeV are presented at both LO and NLO with $m_c=1.5$ GeV and $\mu_r=2m_c$.  
%The effect of cut and  no cut are compared at $m_c=1.5$ GeV and $\mu_r=2m_c$. 
At low reduced CM energy there is obvious effect of the angle cut on  NLO results and the effect becomes smaller as the energy becomes larger. 
Meanwhile, LO results with and without cut are almost coincide, which means the effect of cut is negligible. 
Besides, in comparison with  LO results,  the NLO result shows larger contribution from  NLO correction  at lower reduced CM energy, and the correction becomes smaller as the reduced CM energy becomes closer to initial CM energy. This behavior of  NLO correction consists with what we have seen in Fig.~\ref{total3}. 
\begin{figure}
\centering
\includegraphics[width=8cm]{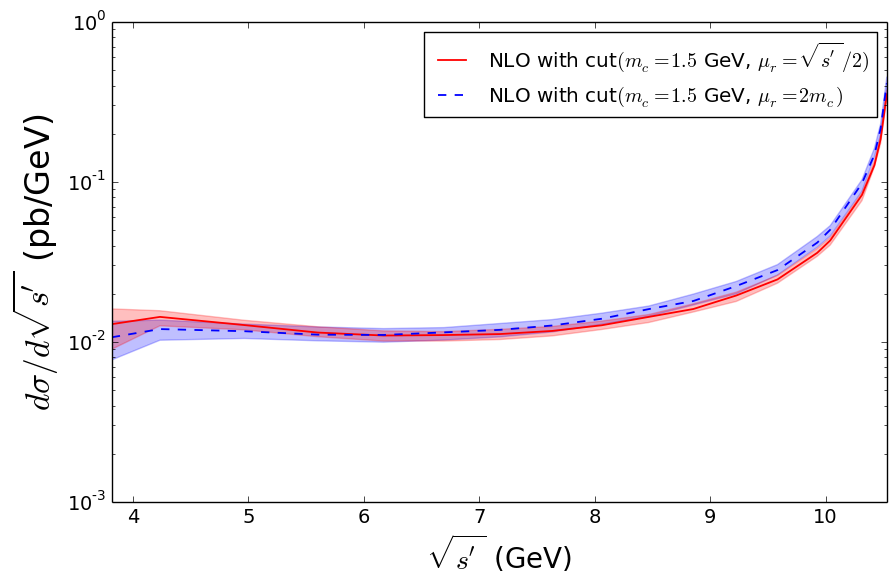}\\
\includegraphics[width=8cm]{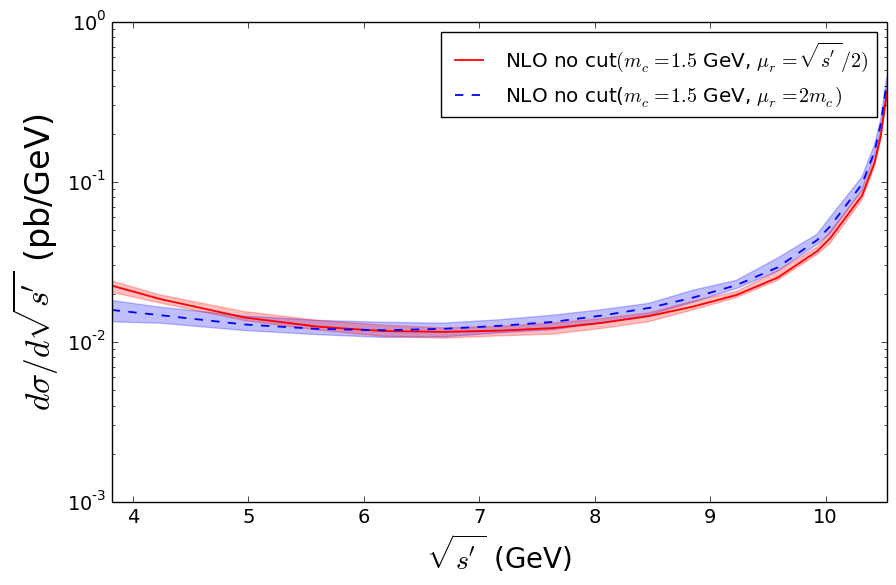}
\caption{\label{fig:ISR3} ISR effect for $J/\psi+gg$ at QCD NLO with the error.}
\end{figure}
The error caused by the uncertainty of charm mass and  the renormalization scale is shown in  Fig.~\ref{fig:ISR3}. 
Each center curve which describes the behavior of $m_c=1.5$ GeV is followed by a band that the upper and lower boundaries are corresponding to $m_c=1.4$ and $1.6$ GeV.
From Fig.~\ref{total3}, we know clearly 
that the measurement for $J/\psi \pi^+ \pi^-$ by BESIII is a few times larger than 
the QCD NLO prediction for inclusive $J/\psi +X$ production, 
therefore experimental measurement for $J/\psi \pi^+ \pi^-$ via ISR effect will certainly be larger than the curve in Fig.~\ref{fig:ISR3}
for the reduced CM energy ranging from 3.70 to 4.60 GeV. 
It means that the curve of ISR effect measured in future experiment will break QCD perturbative prediction
in the small reduced CM energy range but be in agreement with QCD perturbative prediction at large reduced CM energy range. So the comparison between experimental measurement and QCD perturbative prediction in the curve is expecting.  

KEKB has a peak luminosity of 2.1083$\times 10^{34}cm^{-2}s^{-1}$ and the
 SuperKEKB project requires a  peak luminosity of $8\times 10^{35}cm^{-2}s^{-1}$ which is 40 times larger than  KEKB.
SuperKEKB is designed to operate for 11 years. 
The total integrated luminosity accumulated by  the Belle detector have reached $1.04{\rm ab}^{-1}$\cite{Abe:2013kxa} . 
The goal of  the BelleII detector in SuperKEKB is to accumulate an integrated luminosity of $50{\rm ab}^{-1}$\cite{Akai:2018mbz}.
In Table \ref{tab:2}, with the parameter set by $m_c=1.5$ GeV and $\mu_r=2m_c$, the number of events are estimated roughly 
at different bins of reduced CM energy. 
\begin{table*}
\begin{center}
%\footnotesize
\begin{tabular}{|c|c|c|c|c|c|c|}
\hline
reduced CM energy $\sqrt{s^\prime}$(GeV)  & $3.8-6.5 $ & $6.5-8.5 $  & $8.5-9.5$ & $9.5-10.0 $ &$10.0-10.5 $&$10.5-10.583 $\\
\hline
\hline
$\sigma$ without cut (pb)& 0.045  & 0.034 &0.026  &  0.022 &0.060 &0.254\\  
\hline     
Number at KEKB ($\times 10^4$) & 4.70&3.51 & 2.71  &  2.29 &6.26 &26.42\\  
\hline                                                                                             
Number at SuperKEKB($\times 10^6$) & 2.26&1.69 &1.30  &  1.11 &3.01 &12.70\\ 
\hline                         
\hline
$\sigma$ with cut (pb)& 0.044  & 0.033 &0.026  &  0.022 & 0.060 &0.254 \\  
\hline     
Number at KEKB($\times 10^4$) & 4.60&3.48 & 2.69  &  2.29 &6.21 &26.42 \\  
\hline                                                                                             
Number at SuperKEKB($\times 10^6$) & 2.21&1.67 &1.29  &  1.10 &2.99 &12.70 \\  
\hline                                                                                
\end{tabular}
\end{center}
\caption{\label{tab:2} Total cross sections  at different bins of reduced CM energy and the number of events estimated for the designed SuperKEKB experiments of $J/\psi+gg$.}
\end{table*}

\begin{figure}
\begin{center}
\includegraphics[width=8cm]{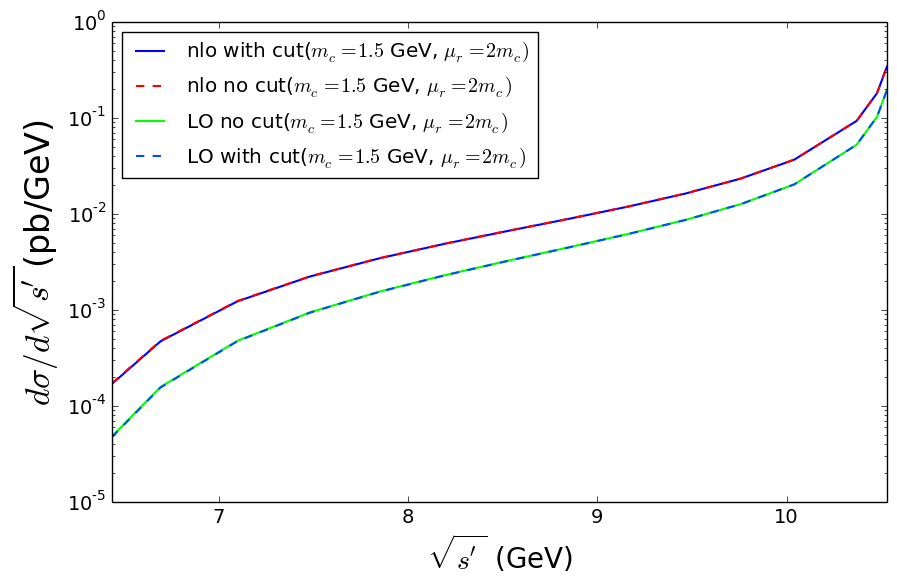}
\caption{\label{fig:ISR5} ISR effect for $J/\psi+c\bar{c}$ at QCD LO and NLO with and without cut.}
\end{center}
\end{figure}

\begin{figure}
\begin{center}
\includegraphics[width=8cm]{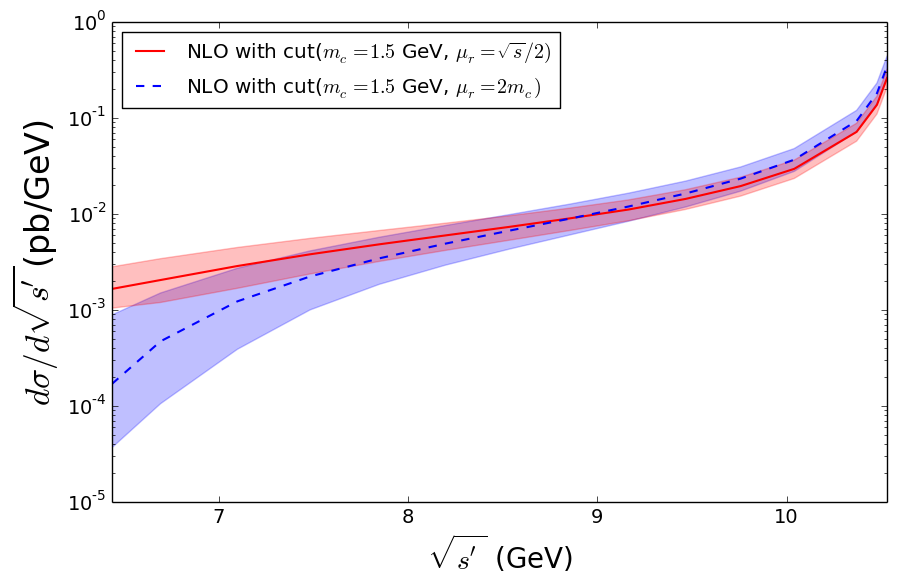} \\
\includegraphics[width=8cm]{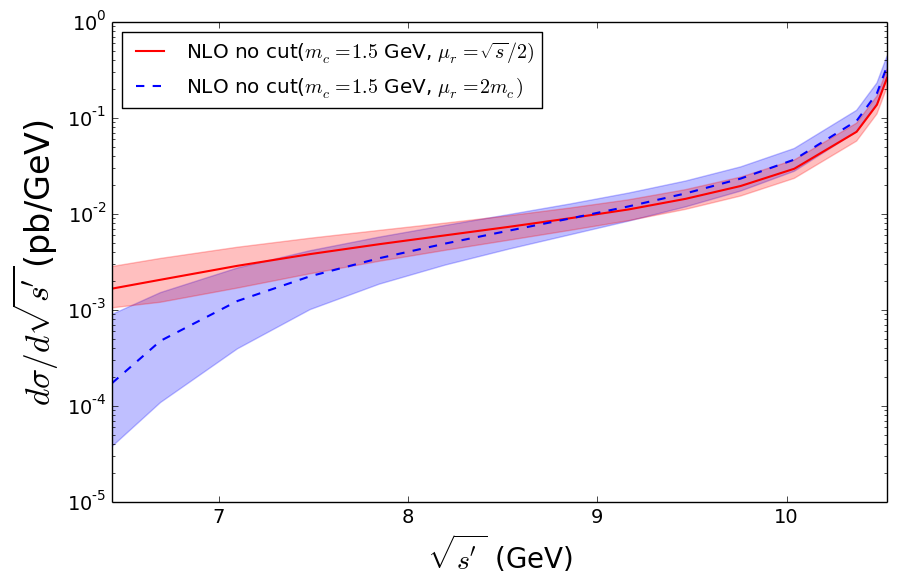}
\caption{\label{fig:ISR8} ISR effect for $J/\psi+c\bar{c}$ at QCD NLO with the error.}
\end{center}
\end{figure}

Similar calculations in $e^+e^-\to J/\psi c\bar{c}$ is also performed.
The cut effect is presented in Fig.~\ref{fig:ISR5}. It shows  that the effect of the angle cut in the detector  is less than 0.1\% no matter at LO or NLO, 
and the two curves are almost overlap with each other.
The same situation is also seen in  Fig.~\ref{fig:ISR8}. 
From  Fig.~\ref{fig:ISR1} and Fig.~\ref{fig:ISR5}, the results are consistent with the ones in Refs.~\cite{Gong:2009kp,Gong:2009ng} when $\sqrt{s^\prime}$ close to 10.58 GeV.
And the numbers of events estimated for  the SuperKEKB  are shown in Table~\ref{tab:3}.
\begin{table*}
\begin{center}
%\footnotesize
\begin{tabular}{|c|c|c|c|c|c|}
\hline
reduce CM energy $\sqrt{s^\prime}$(GeV)   & $6.4-8.5 $  & $8.5-9.5$ & $9.5-10.0 $ &$10.0-10.5 $&$10.5-10.583 $\\
\hline
\hline
$\sigma$ without cut (pb) & 0.006 &0.011  & 0.012 &0.038 &0.169 \\  
\hline     
Number at KEKB($\times 10^4$)&0.62 & 1.17  &  1.27 &3.92 &17.58\\  
\hline                                                                                          
Number at SuperKEKB($\times 10^6$)&0.30 & 0.56  &  0.61 &1.88 &8.45 \\         
\hline
\hline
$\sigma$ with cut (pb)  & 0.006 &0.011  & 0.012 &0.038 &0.169 \\  
\hline    
Number at KEKB($\times 10^4$)&0.62 &1.16 &  1.27&3.90 &17.58 \\  
\hline                                                                                          
Number at SuperKEKB($\times 10^6$)&0.30 &0.56  &  0.61 &1.88 &74.62 \\                                                                                               
\hline
\end{tabular}
\caption{\label{tab:3} Total cross sections  at different bins of reduced CM energy and the number of events estimated for  the designed SuperKEKB experiments of $J/\psi+c\bar{c}$.}
\end{center}
\end{table*}

%**********************************************************************************
\section{Summary and Conclusion}
In summary,  the NLO QCD corrections of inclusive $J/\psi$ production in $e^+e^-$ annihilation with $\sqrt{s}$ range from 3.7 to 10.58 GeV are calculated. And it is found that even the QCD NLO results of the CM energy from 3.7 to 4.6 GeV are still far away from the recent experimental measurements of the BESIII. The perturbative prediction becomes bad when the CM energy is closer to $J/\psi$ production threshold. On the other side close to 10.58 GeV, the result is consistent with the measurements in the Belle\cite{Pakhlov:2009nj}.
It is interesting to study the comparison between experimental measurement and QCD perturbative prediction for the processes with the CM energy from 3.7 to 10.6 GeV.   
By utilizing the ISR effect of  QED, 
the ISR effect of $e^+e^-\to J/\psi+gg$ and $e^+e^-\to J/\psi+c \bar{c}$  are calculated at the QCD NLO with consideration of the uncertainty from  the charm quark mass 
and renormalization scale. The results are plotted and the number of event for different reduced CM energy bin are provided for future SuperKEKB. 
It provides a way to precisely test the validity of perturbative prediction on $J/\psi$ production at $e^+e^-$ collider with different reduced CM energy. 
So we suggest to measure the ISR effect on $J/\psi$ production in future experiments.

This work was supported by the National Natural Science Foundation of China with Grant No. 11475183 and the Key Research Program of Frontier Sciences, CAS, Grant No. Y7292610K1.

%\bigskip

\bibliographystyle{unsrt}
\bibliography{paper}

\begin{thebibliography}{10}

\bibitem{Gross:1973id}
David~J. Gross and Frank Wilczek.
\newblock {Ultraviolet Behavior of Nonabelian Gauge Theories}.
\newblock {\em Phys. Rev. Lett.}, 30:1343--1346, 1973.
\newblock [,271(1973)].

\bibitem{Politzer:1973fx}
H.~David Politzer.
\newblock {Reliable Perturbative Results for Strong Interactions?}
\newblock {\em Phys. Rev. Lett.}, 30:1346--1349, 1973.
\newblock [,274(1973)].

\bibitem{Abe:1992ww}
F.~Abe et~al.
\newblock Inclusive $j/\psi$, $\psi(2s)$ and $b$ quark production in $\bar{p}p$
  collisions at $\sqrt{s} = 1.8$ tev.
\newblock {\em Phys. Rev. Lett.}, 69:3704--3708, 1992.

\bibitem{Bodwin:1994jh}
Geoffrey~T. Bodwin, Eric Braaten, and G.~Peter Lepage.
\newblock Rigorous qcd analysis of inclusive annihilation and production of
  heavy quarkonium.
\newblock {\em Phys. Rev.}, D51:1125--1171, 1995.

\bibitem{Khachatryan:2010yr}
Vardan Khachatryan et~al.
\newblock {Prompt and non-prompt J/psi production in pp collisions at sqrt(s) =
  7 TeV}.
\newblock {\em Eur.Phys.J.}, C71:1575, 2011.

\bibitem{Aaij:2013nlm}
R~Aaij et~al.
\newblock {Measurement of $J/\psi$ polarization in $pp$ collisions at
  $\sqrt{s}=7$ TeV}.
\newblock {\em Eur. Phys. J.}, C73(11):2631, 2013.

\bibitem{Aaij:2014qea}
Roel Aaij et~al.
\newblock {Measurement of $\psi(2S)$ polarisation in $pp$ collisions at
  $\sqrt{s}=7$ TeV}.
\newblock {\em Eur. Phys. J.}, C74(5):2872, 2014.

\bibitem{Campbell:2007ws}
J.~Campbell, F.~Maltoni, and F.~Tramontano.
\newblock Qcd corrections to j/psi and upsilon production at hadron colliders.
\newblock {\em Phys. Rev. Lett.}, 98:252002, 2007.

\bibitem{Gong:2008sn}
Bin Gong and Jian-Xiong Wang.
\newblock {QCD corrections to J/psi polarization of hadronproduction at
  Tevatron and LHC}.
\newblock {\em Phys. Rev. Lett.}, 100:232001, 2008.

\bibitem{Gong:2008ft}
Bin Gong, Xue~Qian Li, and Jian-Xiong Wang.
\newblock {QCD corrections to $J/\psi$ production via color octet states at
  Tevatron and LHC}.
\newblock {\em Phys. Lett.}, B673:197--200, 2009.

\bibitem{Butenschoen:2010rq}
Mathias Butenschoen and Bernd~A. Kniehl.
\newblock {Reconciling J/psi production at HERA, RHIC, Tevatron, and LHC with
  NRQCD factorization at next-to-leading order}.
\newblock {\em Phys. Rev. Lett.}, 106:022003, 2011.

\bibitem{Ma:2010yw}
Yan-Qing Ma, Kai Wang, and Kuang-Ta Chao.
\newblock {J/psi (psi') production at the Tevatron and LHC at
  O($\alpha_s^4v^4$) in nonrelativistic QCD}.
\newblock {\em Phys. Rev. Lett.}, 106:042002, 2011.

\bibitem{Butenschoen:2011yh}
Mathias Butenschoen and Bernd~A. Kniehl.
\newblock {World data of J/psi production consolidate NRQCD factorization at
  NLO}.
\newblock {\em Phys.Rev.}, D84:051501, 2011.

\bibitem{Ma:2010jj}
Yan-Qing Ma, Kai Wang, and Kuang-Ta Chao.
\newblock {A complete NLO calculation of the $J/\psi$ and $\psi'$ production at
  hadron colliders}.
\newblock {\em Phys. Rev.}, D84:114001, 2011.

\bibitem{Butenschoen:2012px}
Mathias Butenschoen and Bernd~A. Kniehl.
\newblock {J/psi polarization at Tevatron and LHC: Nonrelativistic-QCD
  factorization at the crossroads}.
\newblock {\em Phys.Rev.Lett.}, 108:172002, 2012.

\bibitem{Chao:2012iv}
Kuang-Ta Chao, Yan-Qing Ma, Hua-Sheng Shao, Kai Wang, and Yu-Jie Zhang.
\newblock {$J/\psi$ Polarization at Hadron Colliders in Nonrelativistic QCD}.
\newblock {\em Phys. Rev. Lett.}, 108:242004, 2012.

\bibitem{Gong:2012ug}
Bin Gong, Lu-Ping Wan, Jian-Xiong Wang, and Hong-Fei Zhang.
\newblock {Polarization for Prompt $J/\psi$ and $\Psi(2s)$ Production at the
  Tevatron and LHC}.
\newblock {\em Phys. Rev. Lett.}, 110(4):042002, 2013.

\bibitem{Feng:2018ukp}
Yu~Feng, Bin Gong, Chao-Hsi Chang, and Jian-Xiong Wang.
\newblock {The remaining parts for the long-standing $J/\psi$ polarization
  puzzle}.
\newblock 2018.

\bibitem{Braaten:1995ez}
Eric Braaten and Yu-Qi Chen.
\newblock {Signature for color octet production of J / Psi in e+ e-
  annihilation}.
\newblock {\em Phys. Rev. Lett.}, 76:730--733, 1996.

\bibitem{Aubert:2001pd}
B.~Aubert et~al.
\newblock {Measurement of $J/\psi$ production in continuum $e^+e^-$
  annihilations near $\sqrt{s}=10.6$ GeV}.
\newblock {\em Phys. Rev. Lett.}, 87:162002, 2001.

\bibitem{Abe:2001za}
K.~Abe et~al.
\newblock {Production of prompt charmonia in e+ e- annihilation at s**(1/2) =
  10.6-GeV}.
\newblock {\em Phys. Rev. Lett.}, 88:052001, 2002.

\bibitem{Abe:2002rb}
K.~Abe et~al.
\newblock Observation of double c anti-c production in e+ e- annihilation at
  s**(1/2) approx. 10.6-gev.
\newblock {\em Phys. Rev. Lett.}, 89:142001, 2002.

\bibitem{Aubert:2005tj}
B.~Aubert et~al.
\newblock Measurement of double charmonium production in $e^+e^-$ annihilations
  at $\sqrt{s}=10.6$ gev.
\newblock {\em Phys. Rev.}, D72:031101, 2005.

\bibitem{Fleming:2003gt}
Sean Fleming, Adam~K. Leibovich, and Thomas Mehen.
\newblock {Resumming the color-octet contribution to e+ e- --> J/psi + X}.
\newblock {\em Phys. Rev.}, D68:094011, 2003.

\bibitem{Wang:2003fw}
Jian-Xiong Wang.
\newblock {The color octet effect from e+ e- --> J/psi + X + gamma at B
  factory}.
\newblock 2003.

\bibitem{Keung:1980ev}
Wai-Yee Keung.
\newblock {OFF RESONANCE PRODUCTION OF HEAVY VECTOR QUARKONIUM STATES IN e+ e-
  ANNIHILATION}.
\newblock {\em Phys. Rev.}, D23:2072, 1981.

\bibitem{Liu:2002wq}
Kui-Yong Liu, Zhi-Guo He, and Kuang-Ta Chao.
\newblock Problems of double charm production in e+ e- annihilation at s**(1/2)
  = 10.6-gev. ((v)).
\newblock {\em Phys. Lett.}, B557:45--54, 2003.

\bibitem{Zhang:2006ay}
Yu-Jie Zhang and Kuang-Ta Chao.
\newblock Double charm production e+ e- --> j/psi + c anti-c at b factories
  with next-to-leading order qcd correction.
\newblock {\em Phys. Rev. Lett.}, 98:092003, 2007.

\bibitem{Ma:2008gq}
Yan-Qing Ma, Yu-Jie Zhang, and Kuang-Ta Chao.
\newblock {QCD correction to $\bm{e^+ e^- \to J/\psi g g}$ at B Factories}.
\newblock {\em Phys. Rev. Lett.}, 102:162002, 2009.

\bibitem{Gong:2009kp}
Bin Gong and Jian-Xiong Wang.
\newblock {Next-to-Leading-Order QCD Corrections to $e^ + e^- -> J/\psi+gg$ at
  the B Factories}.
\newblock {\em Phys. Rev. Lett.}, 102:162003, 2009.

\bibitem{Gong:2009ng}
Bin Gong and Jian-Xiong Wang.
\newblock {Next-to-leading-order QCD corrections to $e^+e^- \to J/\psi_{cc} $
  at the $B$ factories}.
\newblock {\em Phys. Rev.}, D80:054015, 2009.

\bibitem{Zhang:2009ym}
Yu-Jie Zhang, Yan-Qing Ma, Kai Wang, and Kuang-Ta Chao.
\newblock {QCD radiative correction to color-octet J/psi inclusive production
  at B Factories}.
\newblock {\em Phys.Rev.}, D81:034015, 2010.

\bibitem{Ablikim:2016qzw}
Medina Ablikim et~al.
\newblock {Precise measurement of the $e^+e^-\to \pi^+\pi^-J/\psi$ cross
  section at center-of-mass energies from 3.77 to 4.60 GeV}.
\newblock {\em Phys. Rev. Lett.}, 118(9):092001, 2017.

\bibitem{Li:2014fya}
Yi-Jie Li, Guang-Zhi Xu, Pan-Pan Zhang, Yu-Jie Zhang, and Kui-Yong Liu.
\newblock {Study of Color Octet Matrix Elements Through $J/\psi$ Production in
  $e^{+}e^{-}$ Annihilation}.
\newblock {\em Eur. Phys. J.}, C77(9):597, 2017.

\bibitem{FDC}
Jian-Xiong Wang.
\newblock Progress in fdc project.
\newblock {\em Nucl. Instrum. Meth.}, A534:241--245, 2004.

\bibitem{Harris:2001sx}
B.~W. Harris and J.~F. Owens.
\newblock The two cutoff phase space slicing method.
\newblock {\em Phys. Rev.}, D65:094032, 2002.

\bibitem{Patrignani:2016xqp}
C.~Patrignani et~al.
\newblock {Review of Particle Physics}.
\newblock {\em Chin. Phys.}, C40(10):100001, 2016.

\bibitem{Pakhlov:2009nj}
P.~Pakhlov et~al.
\newblock {Measurement of the $e^+e^-\to J/\psi c \bar{c}$ cross section at
  s**(1/2) $\approx$ 10.6 GeV}.
\newblock {\em Phys.Rev.}, D79:071101, 2009.

\bibitem{Yuan:2007sj} 
  C.~Z.~Yuan {\it et al.} [Belle Collaboration],
  %``Measurement of e+ e- ---> pi+ pi- J/psi cross-section via initial state radiation at Belle,''
  Phys.\ Rev.\ Lett.\  {\bf 99}, 182004 (2007)
  doi:10.1103/PhysRevLett.99.182004
  [arXiv:0707.2541 [hep-ex]].
  %%CITATION = doi:10.1103/PhysRevLett.99.182004;%%
  %411 citations counted in INSPIRE as of 05 Mar 2019

\bibitem{Wang:2007ea} 
  X.~L.~Wang {\it et al.} [Belle Collaboration],
  %``Observation of Two Resonant Structures in e+e- to pi+ pi- psi(2S) via Initial State Radiation at Belle,''
  Phys.\ Rev.\ Lett.\  {\bf 99}, 142002 (2007)
  doi:10.1103/PhysRevLett.99.142002
  [arXiv:0707.3699 [hep-ex]].
  %%CITATION = doi:10.1103/PhysRevLett.99.142002;%%
  %397 citations counted in INSPIRE as of 05 Mar 2019 

\bibitem{Chang:2010am} 
  C.~H.~Chang, J.~X.~Wang and X.~G.~Wu,
  %``Production of a Heavy Quarkonium with a Photon or via ISR at $Z$ Peak in $e^+e^-$ Collider,''
  Sci.\ China Phys.\ Mech.\ Astron.\  {\bf 53}, 2031 (2010)
  doi:10.1007/s11433-010-4142-7
  [arXiv:1005.4723 [hep-ph]].
  %%CITATION = doi:10.1007/s11433-010-4142-7;%%
  %11 citations counted in INSPIRE as of 19 Mar 2019


\bibitem{Lee:1964is}
T.~D. Lee and M.~Nauenberg.
\newblock {Degenerate Systems and Mass Singularities}.
\newblock {\em Phys. Rev.}, 133:B1549--B1562, 1964.
\newblock [,25(1964)].

\bibitem{Kuraev:1985hb}
E.~A. Kuraev and Victor~S. Fadin.
\newblock {On Radiative Corrections to e+ e- Single Photon Annihilation at
  High-Energy}.
\newblock {\em Sov. J. Nucl. Phys.}, 41:466--472, 1985.
\newblock [Yad. Fiz.41,733(1985)].

\bibitem{Nicrosini:1986sm}
O.~Nicrosini and Luca Trentadue.
\newblock {Soft Photons and Second Order Radiative Corrections to e+ e- --->
  Z0}.
\newblock {\em Phys. Lett.}, B196:551, 1987.

\bibitem{Greco:2016izi}
Mario Greco, Tao Han, and Zhen Liu.
\newblock {ISR effects for resonant Higgs production at future lepton
  colliders}.
\newblock {\em Phys. Lett.}, B763:409--415, 2016.

\bibitem{Kuraev:1985wb}
E.~A. Kuraev and Victor~S. Fadin.
\newblock {CALCULATION OF RADIATIVE CORRECTIONS TO THE CROSS-SECTION OF ONE
  PHOTON ANNIHILATION BY MEANS OF STRUCTURE FUNCTIONS}.
\newblock 1985.

\bibitem{Akai:2018mbz}
Kazunori Akai, Kazuro Furukawa, and Haruyo Koiso.
\newblock {SuperKEKB Collider}.
\newblock {\em Nucl. Instrum. Meth.}, A907:188--199, 2018.

\bibitem{Lalwani:2018dgg}
Kavita Lalwani and Manish Kumar.
\newblock {Belle II Experiment: Status and Upgrade}.
\newblock {\em Few Body Syst.}, 59(6):142, 2018.

\bibitem{Abe:2013kxa}
Tetsuo Abe et~al.
\newblock {Achievements of KEKB}.
\newblock {\em PTEP}, 2013:03A001, 2013.

\end{thebibliography}
%\begin{thebibliography}{90}
%
%\vspace{3mm}
%%
%%\bibitem{lab1}M. L. Liu, Y. H. Zhang, X. H. Zhou et al, Phys. Rev. C, {\bf 70}: 14---34 (2004)
%%
%%\bibitem{lab2} M. Tinkham, \emph{Group Theory and Quantum Mechanics} (New
%%York: McGraw-Hill, 1964), p. 10---50
%%
%%\bibitem{lab3} T. Tel, in \emph{Chaos}, edited by B. Hao (Singapore: World Scientific, 1990), p. 149
%
%\bibitem{NRQCD} G.T.Bodwin,E.Braaten,and G.P.Lepage, 
%%
%Phys.Rev.D51,1125(1995).
%
%
%\bibitem{exp1} M. Ablikim et al. 
%%(BESIII Collaboration), “Precise measurement of the e+e− → π+π−J/ψ cross section at CS energy from 3.77 to 4.60 GeV”, Phys.
%Phys.Rev. Lett. 118, 092001 (2017)
%
%\bibitem{BES3} M. Ablikim et al. 
%%
%(BESIII Collaboration), Chin. Phys.C39, 093001(2015).
%
%\bibitem{ee} B. P. Abbott, et al,
%%Observation of gravitational waves from a binary black hole merger,
%Phys. Rev. Lett. {\bf 116}: 061102 (2016)
%
%\bibitem{fdc} J.X. Wang,
%%
%Nucl.Instrum.Meth.A534,241(2004)
%
%\bibitem{renormalization} B. Gong and J. X. Wang, Phys.Rev. D77, 054028(2008) 
%%
%
%\bibitem{two-cut} B. W. Harris and J. F. Owens. 
%%
%Phys. Rev. D65, 094032(2002).
%
%
%\bibitem{ee2}B. P. Abbott, et al,
%%GW151226: Observation of Gravitational Waves from a 22-Solar-Mass Binary Black Hole Coalescence
%Phys. Rev. Lett. {\bf 116}: 241103 (2016)
%
%
%
%\end{thebibliography}
%\end{multicols}

%\clearpage
%\end{CJK*}
\end{document}